\documentclass[conference]{IEEEtran}
\IEEEoverridecommandlockouts
\usepackage{cite}
\usepackage{amsmath,amssymb,amsfonts}
\usepackage{graphicx}
\usepackage{textcomp}
\usepackage{xcolor}
\def\BibTeX{{\rm B\kern-.05em{\sc i\kern-.025em b}\kern-.08em
    T\kern-.1667em\lower.7ex\hbox{E}\kern-.125emX}}

\usepackage{verbatim}
\usepackage{comment}
\usepackage{algorithm,algpseudocode}
\usepackage{url}

\graphicspath{ {./image/} }

\def \cC{\mathcal{C}}

\makeatletter
\newcommand\fs@spaceruled{\def\@fs@cfont{\bfseries}\let\@fs@capt\floatc@ruled
  \def\@fs@pre{\vspace{1\baselineskip}\hrule height.8pt depth0pt \kern2pt}%
  \def\@fs@post{\kern2pt\hrule\relax}%
  \def\@fs@mid{\kern2pt\hrule\kern2pt}%
  \let\@fs@iftopcapt\iftrue}
\makeatother

\begin{document}

\title{CRC Codes as Error Correction Codes\\
}
%\author{\IEEEauthorblockN{1\textsuperscript{st} Wei An}
\author{\IEEEauthorblockN{Wei An}
\IEEEauthorblockA{\textit{Research Laboratory of Electronics} \\
\textit{Massachusetts Institute of Technology}\\
Cambridge, MA 02139, USA \\
wei\_an@mit.edu}
\and
%\IEEEauthorblockN{2\textsuperscript{nd} Muriel M\'edard}
\IEEEauthorblockN{Muriel M\'edard}
\IEEEauthorblockA{\textit{Research Laboratory of Electronics} \\
\textit{Massachusetts Institute of Technology}\\
Cambridge, MA 02139, USA \\
medard@mit.edu}
\and
%\IEEEauthorblockN{3\textsuperscript{rd} Ken R. Duffy}
\IEEEauthorblockN{Ken R. Duffy}
\IEEEauthorblockA{\textit{Hamilton Institute} \\
\textit{Maynooth University}\\
Ireland \\
ken.duffy@mu.ie}
}

\maketitle

\begin{abstract}
CRC codes have long since been adopted in a vast range of applications. The established notion that they are suitable primarily for error detection can be set aside through use of the recently proposed Guessing Random Additive Noise Decoding (GRAND). Hard-detection (GRAND-SOS) and soft-detection (ORBGRAND) variants can decode any short, high-rate block code, making them suitable for error correction of CRC-coded data. When decoded with GRAND, short CRC codes have error correction capability that is at least as good as popular codes such as BCH codes, but with no restriction on either code length or rate.

The state-of-the-art CA-Polar codes are concatenated CRC and Polar codes. For error correction, we find that the CRC is a better short code than either Polar or CA-Polar codes. Moreover, the standard CA-SCL decoder only uses the CRC for error detection and therefore suffers severe performance degradation in short, high rate settings when compared with the performance GRAND provides, which uses all of the CA-Polar bits for error correction.

Using GRAND, existing systems can be upgraded from error detection to low-latency error correction without re-engineering the encoder, and additional applications of CRCs can be found in IoT, Ultra-Reliable Low Latency Communication (URLLC), and beyond. The universality of GRAND, its ready parallelized implementation in hardware, and the good performance of CRC as codes make their combination a viable solution for low-latency applications.

\end{abstract}

\begin{IEEEkeywords}
CRC; Polar; Error Correction; GRAND; URLLC.
\end{IEEEkeywords}

\section{Introduction} \label{sect:intro}
Owing to the simplicity of their implementation and their powerful
error detection capability, Cyclic Redundancy Checks (CRCs) \cite{crc1961}
are widely adopted in applications in storage, networking, and
communications.  CRCs are sometimes used not only to error detect, but also to assist in error correction. For
example, to assist in early stopping of Turbo decoding iterations
\cite{4656995}, or joint decoding of CRCs and convolutional codes
\cite{4686263}.  In list decoding architectures, a CRC is commonly
used for selection from a list of candidates. This technique has been
applied to Turbo codes \cite{list_turbo} and Polar codes
(CA-Polar) \cite{Arikan09,KK-CA-Polar,TV-list}. The latter leads
to the success of CRC-aided successive cancellation list (CA-SCL
or SCL for brief) decoding algorithms. The better performance at shorter code lengths of
 CA-Polar codes, as compared to other common Shannon limit
approaching codes, such as Turbo codes and LDPC
codes \cite{ldpc97}, has led to their inclusion in 5G-NR control communications where latency is of particular
concern \cite{3gpp38212}.

With the error correction capability enabled by CRC, significant
benefits can be brought to existing systems, but have heretofore
seen only limited use. For example, the Automatic Repeat Request
(ARQ) scheme of networking can be upgraded to Hybrid ARQ through the
use of CRC's error correction. Similar benefits can be found in IoT or personal area network (PAN). Previous research on the use of CRCs for error correction focused only on fixing one or two erroneous bits \cite{crc_fec_2006, crc2020}. Soft detection techniques such as Belief Propagation (BP) and Linear Programming (LP) \cite{crcLLR} have also been considered for decoding CRCs, but their performance is severely limited when used for short and high rate codes. No previous approach leads to a broadly applicable decoding method capable of extracting the maximum error correction performance from CRCs.

Many emerging applications require low latency communication of
small packets of data. Examples include machine communications in
IoT systems \cite{ieee_latency, iot2019, wHP_fec}, telemetry,
tracking, and command in satellite communications \cite{ttc_2005,
fec_ttc_2014}, all control channels in mobile communication systems,
and Ultra-Reliable Low Latency Communications (URLLC) as proposed
in 5G-NR \cite{urllc,sybis2016channel,shirvanimoghaddam2018short}.
Most codes and decoders were designed to function well for large blocks of
data, but provide underwhelming performance for short codes.  As a
result, some conventional codes have received renewed attention
\cite{short_fec, bch_m2m, fec_wireless}, including Reed-Solomon
Codes\cite{ReedSolo}, BCH codes\cite{journals/iandc/BoseR60a} and
Random Linear Codes (RLC). In this paper, we show that CRCs stand
out as a viable class of error correcting codes.

The error correction capability of CRCs is enabled by the recently
introduced Guessing Random Additive Noise Decoding (GRAND) algorithm
\cite{Duffy18,kM:grand}, which can decode any short, high-rate block
codes. GRAND's universality stems from its effort to identify the
effect of the noise, from which the code-word is deduced. Originally
introduced as a hard detection decoder \cite{Duffy18,kM:grand,grand-mo},
a series of soft detection variants,  SRGRAND \cite{Duffy19a,Ken:5G},
ORBGRAND \cite{Duffy20} and SGRAND \cite{Solomon20}, have since been proposed that
make distinct quantization assumptions. Both hard and
soft detection versions of GRAND are  ML decoders. The simplicity of GRAND's operation has
resulted in the proposal of efficient circuit implementations
\cite{abbas2020high}.

With GRAND and standard decoders if available, we evaluate CRCs
as short, high-rate error correction codes when compared with BCH,
RLCs and CA-Polar codes. Using the best published  CRC generator
polynomials \cite{koopmanWeb} we find that CRCs perform as well as
BCH codes at their available settings, but CRCs can operate for a
wider range of code lengths and rates. Featuring even more flexible
code-book settings, RLCs possess security potentials by code-book
re-randomisation, but at very high code-rates RLC performance
degrades in comparison to select CRCs. Using GRAND to compare the
performance of Polar, CA-Polar and CRC codes, we find that that the
CRC is not aiding so much as dominating the decoding performance
potential. Furthermore, the celebrated CA-SCL algorithm underperforms
for short packets because the CRC bits are only used for error
detection.

The rest of the paper is organized as follows. Section \ref{sect:grand}
introduces GRAND hard and soft detection variants. Section
\ref{sect:code} review CRC and other comparison candidates. Section
\ref{sect:sim_comp} provides simulated performance evaluation of
involved channel codes. Section \ref{sect:conclusion} summarizes
the paper's findings.

\section{Guessing Random Additive Noise Decoding} \label{sect:grand}
%GRAND algorithms are outlined here, including GRAND-SOS, a version for hard detection and ORBGRAND, a version for soft detection.
\subsection{GRAND-SOS for hard detection} \label{sub:grand}
Consider a transmitted binary code-word $X^n\in {\cC}$ drawn from an arbitrary rate $R$ code-book  ${\cC}$, i.e. a set of $2^{nR}=2^K$ strings in $\{0,1\}^n$, where  $(n,K)$ is a core pair of parameters of a codebook. Assume independent post-hard-detection channel noise, $N^n$, which also takes values from $\{0,1\}^n$, additively alters $X^n$ between transmission and reception. The resulting sequence is $Y^n = X^n \oplus N^n$, where $\oplus$ represents addition modulo 2.

From $Y^n$ GRAND attempts to determine $X^n$ indirectly by identifying
$N^n$ through sequentially taking putative noise sequences, $z^n$,
which we sometimes term patterns, subtracting them from the received
signal and querying if what remains, $Y^n\oplus z^n$, is in the
code-book $\cC$. If transmitted code-words are all equally likely
and $z^n$ are queried in order from most likely to least likely
based on the true channel statistics, the first instance where a
code-book element is found is an optimally accurate maximum likelihood
decoding in the hard detection setting \cite{kM:grand}. For the
hard detection binary symmetric channels (BSC) channels considered
in this paper the order of testing noise patterns $z^{n,i},
i=1,2,...,n$ follows their Hamming weight from low to high. For
patterns with identical Hamming weights, we choose a sweeping order
of generation, in which a grouped pattern sweeps from one end of
sequence to the other end. A new grouped pattern is obtained by
permuting the bits in the group. When permutation exhausts, an extra
non-flipped bit is included and the permutation process restarts.
This pattern generator gives the decoder some advantages with burst
errors (specifically treated in \cite{grand-mo}) and the resulted algorithm is named as GRAND with simple
order sweeping (GRAND-SOS).

It is established in \cite{kM:grand} that one need not query all
possible noise patterns for the decoder to be capacity-achieving,
and instead one can determine a threshold for the number of code-book
queries, termed the abandonment threshold, at which a failure to
decode can be reported without unduly impacting error correction
performance.  Standard decoders of linear block codes can decode
up to $t=\left \lfloor d/2 \right \rfloor$ errors where $d$ is the
minimum distance of the code-book \cite{lin2004error}. Using this
limit as abandonment threshold, GRAND is guaranteed to achieve at
least equivalent performance to existing code-book dedicated decoders.

\subsection{ORBGRAND for soft detection} \label{sub:orb}
Ordered reliability bits GRAND (ORBGRAND) is a soft detection
approximate ML decoder that features significant complexity advantage
over the full ML soft GRAND (SGRAND) decoder \cite{Solomon20}, and
still provides better BLER performance than state-of-the-art soft
decoders \cite{Duffy20} for short, high-rate codes.

Soft detection versions of GRAND require reliability information
of each received code-word bit. With binary phase-shift key (BPSK) modulation and additive
white Gaussian noise (AWGN) channels, the reliability is simply the
absolute value of the received signal, from which the probability
of a test pattern can be computed and used for its testing order.
Facing the fact that efficient implementation of this simple idea
is not computational straight-forward, ORBGRAND develops an efficient
method to generate a approximate optimal order that is universal.

Assume the reliability order index of the $k-th$ received bit is $r_k^n \in (1,2,...,n)$. The vector $r^n$ records the reliability orders of all received bits in $Y^n$. A reliability weighted Hamming weight called \textit{Logistic Weight} for $z^n$ is defined as,
\begin{equation} \label{eq:lw}
w_L(z^n) = \sum_{k=1}^{n}r_k^nz_k^n
\end{equation}
This metric is used to determine the checking order of testing noise sequences $z^{n,i}, i=1,2,...,n$, i.e. $w_L(z^{n,i})<w_L(z^{n,j})$ for $i<j$. Patterns with identical Logistic Weights can be arbitrarily ordered. 

The performance of soft decoders is no longer restricted by minimum
distance of code-books. %While looser abandonment conditions can
%only lead to performance improvement, 
In most cases ORBGRAND with
a moderate complexity can provide better performance than state-of-the-art
soft decoders.

\section{Overview of Channel Codes} \label{sect:code}

\subsection{CRC codes} \label{sub:crc}
As a type of cyclic code \cite{prange1957cyclic}, CRC operates in
terms of polynomial computation based on Galois field of two elements
GF(2). The CRC generator polynomial can be expressed as
$g(x)=\sum_{k=0}^{n-K-1}g_kx^k$, where $n-K$ is the number of CRC bits
to be appended to a message sequence, expressed as polynomial $m(x)
= \sum_{i=0}^{K-1}m_ix^i$, with $n$ being the code-word length.
The CRC code-word is constructed as,
\begin{equation} \label{eq:crc}
c(x) = x^{n-K}m(x) + \text{remainder}\left ( \frac{x^{n-K}m(x)}{g(x)} \right )
\end{equation}
with the remainder of polynomial division appended to the shifted
message polynomial, the resulting code-word polynomial $c(x)$ is
divisible by $g(x)$. The property is used in parity checks to detect
errors, and provides an efficient implementation for
checking code-book membership in GRAND. While Equation (\ref{eq:crc})
presents the systematic format of CRC codes, they can also be
constructed as other cyclic codes (such as BCH) as,
\begin{equation} \label{eq:crc2}
c(x) = m(x)g(x)
\end{equation}
Equation (\ref{eq:crc2}) generates the same code-book as (\ref{eq:crc}),
but with lower computation complexity.

To maximize the number of detectable errors, the CRC maximizes the
minimum Hamming distance of its code-book. The topic has been studied
extensively and optimal generator polynomials have been published
for a wide set  of combinations of CRC bit numbers and message lengths
\cite{koopman04,koopman06,koopmanWeb}. Unlike channel codes for
error correction function, the design of CRC codes need not consider
decoder implementation, indicating potentially better decoding
performance than normal Forward Error Correction (FEC) codes,
provided that a decoder is available, i.e. GRAND.

\subsection{Other short code candidates} \label{sub:other}
Also as a type of cyclic codes, BCH codes can be constructed with
Equation (\ref{eq:crc2}). In this sense, BCH codes can be viewed as
a special case of CRC codes. For any positive integers $m>3$ and
$t<2^{m-1}$, there exists a binary BCH code with code-word length
of $n=2^m-1$, parity bit number of $n-K \leq mt$ and minimum distance
of $d \geq 2t+1$. A list of available parameters $n$, $K$ and $t$
can be found in \cite{lin2004error}. The standard decoder for BCH
codes is Berlekamp-Massey (B-M) decoder \cite{journals/iandc/BoseR60a},
which is also used for the widely adopted Reed-Solomon (RS) codes,
a special type of BCH codes with operations extended to higher order
Galois fields. It should be noted that the original proposal of
GRAND has no limitation on the Galois field order\cite{kM:grand},
Here we focus on binary codes, but the methodology can be easily extended
to RS codes in higher order Galois fields.

RLCs are linear block codes that have long been used for theoretical investigation,
 but have heretofore not been considered practical in terms
of decodability. With GRAND, RLCs are no more challenging to decode
than any other linear codes. Their generator matrices are obtained
by appending a randomly generated parity component to the identity
matrix, from which the parity check matrix can be computed accordingly.
In addition to its flexibility, the ease of code-book construction
grants RLC  security features by real-time code-book updates.

CA-Polar codes are concatenated CRC and Polar codes. Although they approach the Shannon reliable rate limit \cite{Arikan09}, Polar codes do 
not give satisfactory decoding performance for practical code-word
lengths, even with the ML successive cancellation (SC)
algorithm. A CRC is introduced as an outer code to assist decoding
and the CA-SCL decoding algorithm was invented as the key to the
success of the resulted CA-Polar codes \cite{KK-CA-Polar,TV-list}.
In the decoding process, a list of candidate code-words are produced
and CRC checks are performed to make the selection with confidence. There is to our knowledge no standard hard
detection decoder for Polar or CA-Polar codes apart from
GRAND decoders. For code-book membership checking in GRAND, 
parity check matrices can be easily derived for  generator
matrices, which, for linear block codes, can be obtained with a
simple method that injects each vector of the K-by-K identity matrix
to the encoder and collects the output vectors.

\section{Simulated Performance Evaluation}  \label{sect:sim_comp}
For our simulations, we map the stationary bit flip probability of the
channel energy per transmitted information bit via the usual AWGN
BPSK formula: $
p= \text{Q}\left(\sqrt{2RE_b/N_0}\right),
$
where $R$ is the code rate. Table \ref{tb:bch_crc} lists the code-book
settings considered for CRC evaluation, along with published preferred generator polynomials. Note that the
zero order 1 is omitted in the least significant bit \cite{koopmanWeb}.
\begin{table}[!ht]
\centering
\begin{tabular}{||c c | c | c c||}
 \hline
 N & K & BCH $t$ & CRC poly. & $d$ \\ [0.5ex] 
 \hline
 127 & 120 & 1 & 0x65  & 3 \\ 
 127 & 113 & 2 & 0x212d & 5 \\ 
 127 & 106 & 3 & 0x12faa5 & 7 \\ 
 128 & 99  & N/A  & 0x13a46755  & 8 \\ 
 \hline
 63 & 57 & 1 & 0x33 & 3 \\ 
 63 & 51 & 2 & 0xbae & 5 \\ 
 63 & 45 & 3 & 0x25f6a & 7 \\ 
 64 & 51 & N/A & 0x12e6 & 4 \\ 
 \hline
\end{tabular}
\caption{List of code-book settings along with correctable error number $t$ for BCH codes and minimum distances $d$ for CRC codes.}
\label{tb:bch_crc}
\end{table}

\subsection{CRC Codes vs BCH Codes} \label{sub:crc_bch}
\begin{figure}[htbp]
\centerline{\includegraphics[width=0.44\textwidth]{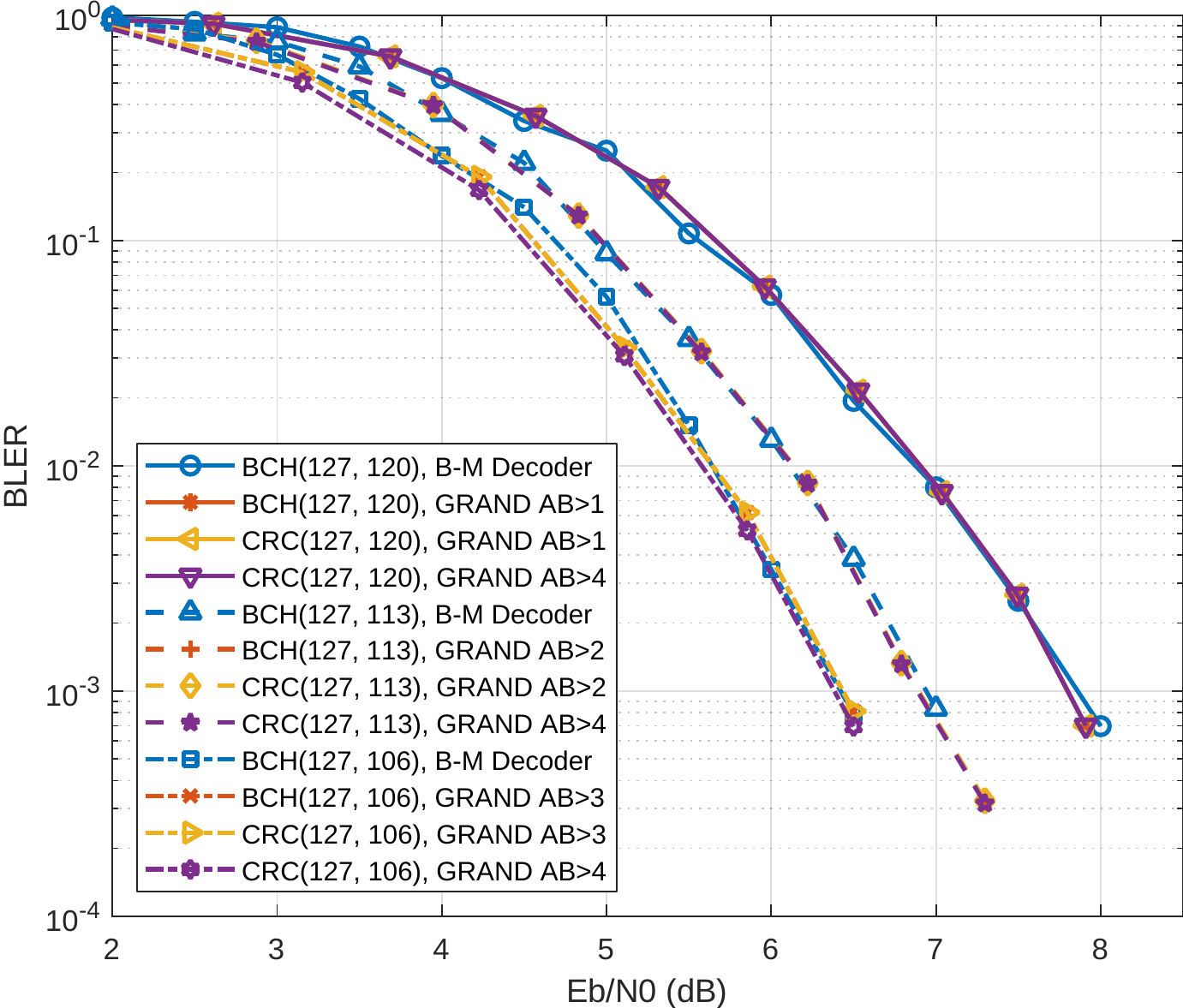}}
\caption{Performance evaluation of code/decoder combinations with code-word length of $127$ for BCH and CRC codes.}
\label{fig:bch_crc_127}
\end{figure}

For code-book settings in Table \ref{tb:bch_crc} available for BCH codes, we measure the performance of the following code/decoder combinations:
\begin{itemize}
  \item BCH codes with Berlekamp-Massey (B-M) decoders
  \item BCH codes with GRAND-SOS decoder and $AB>t$
  \item CRC codes with GRAND-SOS decoder and $AB>t$
  \item CRC codes with GRAND-SOS decoder and $AB>4$
\end{itemize}
$AB>t$ indicates abandonment when the error number in testing patterns is larger than the code-book decoding capability, $t$. And $AB>4$ is a loose condition for all available BCH codes listed in Table \ref{tb:bch_crc}, with which we show the influence of the minimum distance of a code-book to its decoding performance. Simulation results for code-word length of $127$ are presented in Fig. \ref{fig:bch_crc_127}. 

\begin{figure}[htbp]
\centerline{\includegraphics[width=0.44\textwidth]{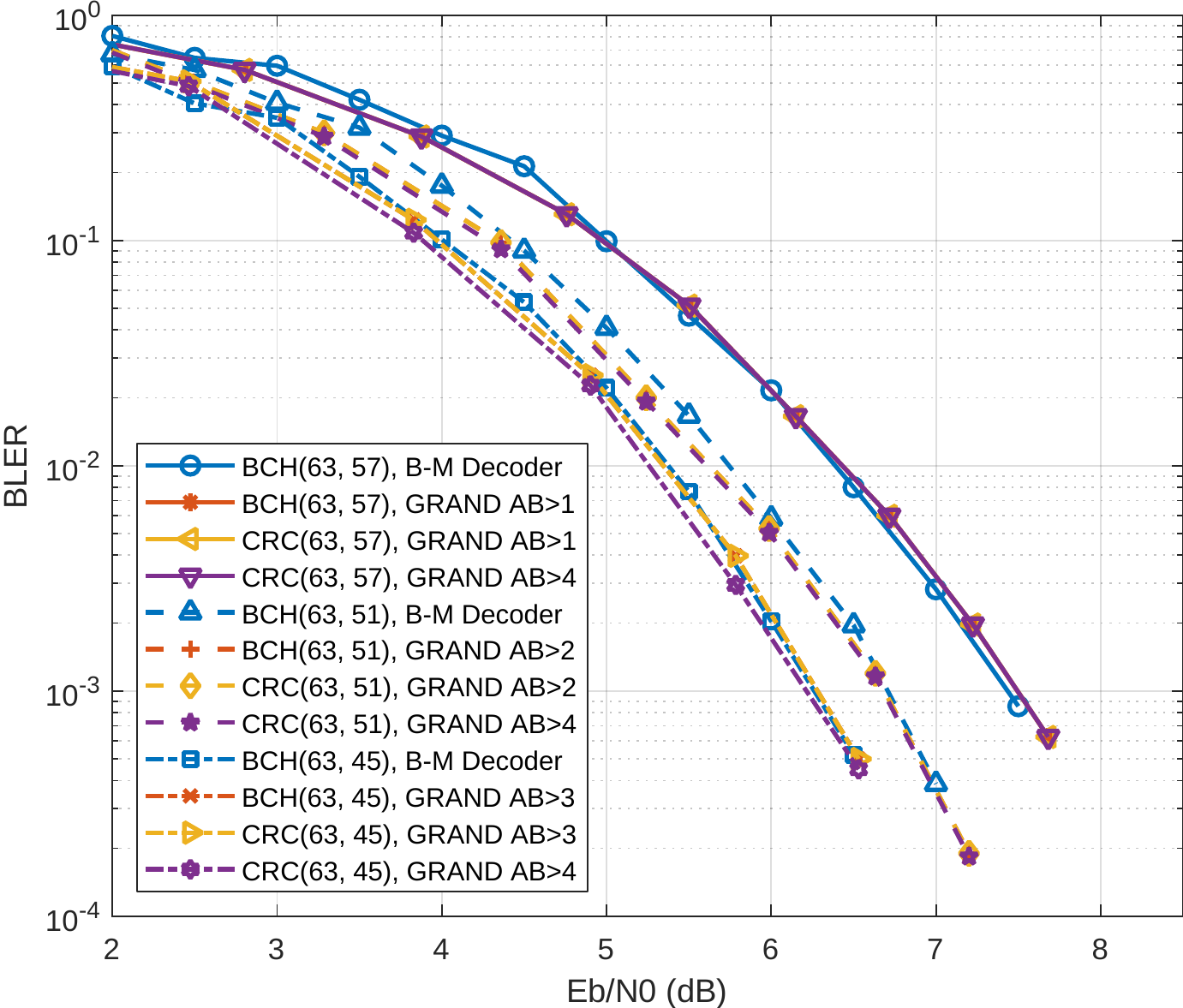}}
\caption{Performance evaluation of code/decoder combinations with
code-word length of $63$ for BCH and CRC codes.} \label{fig:bch_crc_63}
\end{figure}

For each code rate, all four code/decoder combinations have 
performance curves that are close to each other, leading to the following
conclusions. For BCH codes, the GRAND-SOS algorithm has the same decoding
capability as the standard Berlekamp-Massey decoder. With identical
code-book settings, CRC codes have matching performances to BCH
codes. Loosening abandonment condition grants little improvement
to decoding performance of GRAND, so the use of abandonment to curtail complexity remains attractive.
%, allowing the usage of code-book
%minimum distance as the abandonment threshold and the minimized
%GRAND complexity. 
These observations are further confirmed with
simulations of code-word length of $63$, as shown in Fig.
\ref{fig:bch_crc_63} 

Combined evaluation of curves in Fig. \ref{fig:bch_crc_127} and
Fig. \ref{fig:bch_crc_63} indicates that performance of short codes
is sensitive to code-word lengths and rates, therefore the flexibility
of CRC renders it attractive vis-\`a-vis BCH codes in
low-latency scenarios.

\subsection{CRC Codes vs RLCs} \label{sub:crc_rlc}
\begin{figure}[htbp]
\centerline{\includegraphics[width=0.44\textwidth]{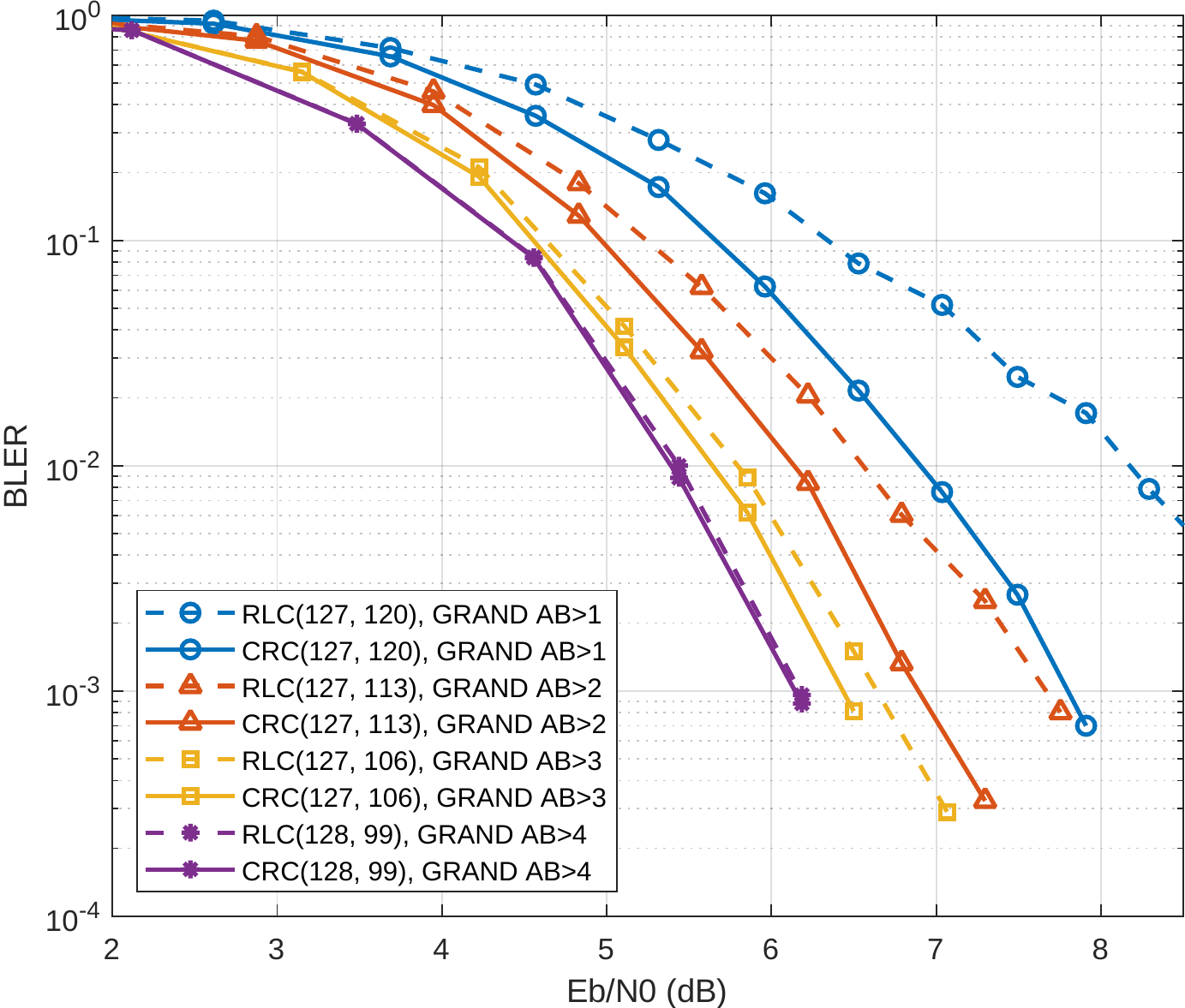}}
\caption{Hard detection performance of RLC and CRC codes with selected code-word lengths and rates.}
\label{fig:rlc_crc}
\end{figure}

To explore fully the decoding capability of RLC codes, we update
the generator matrix for every code-word transmission to achieve
random selection of code-words. For a fair comparison, we apply the
same abandonment threshold for CRC and RLC codes. In Fig.
\ref{fig:rlc_crc}, the RLC(128, 99) curve overlaps the CRC(128, 99)
curve. As the code rate increases, however, the RLC performance
degrades in comparison to the CRC code, owing to the shrinking of
code-word space that makes it more probable for RLC to pick up
code-words close to each other. The generator polynomial of CRC,
however, is the product of a selection procedure
\cite{koopman04,koopman06} that thus results in a code-book with
stable minimum distance. This suggests that to extract near-CRC code
performance for very high rate RLCs, extra structure might be needed.

\subsection{Comparison with Polar and CA-Polar Codes} \label{sub:crc_polar}

\begin{figure}[htbp]
\centerline{\includegraphics[width=0.44\textwidth]{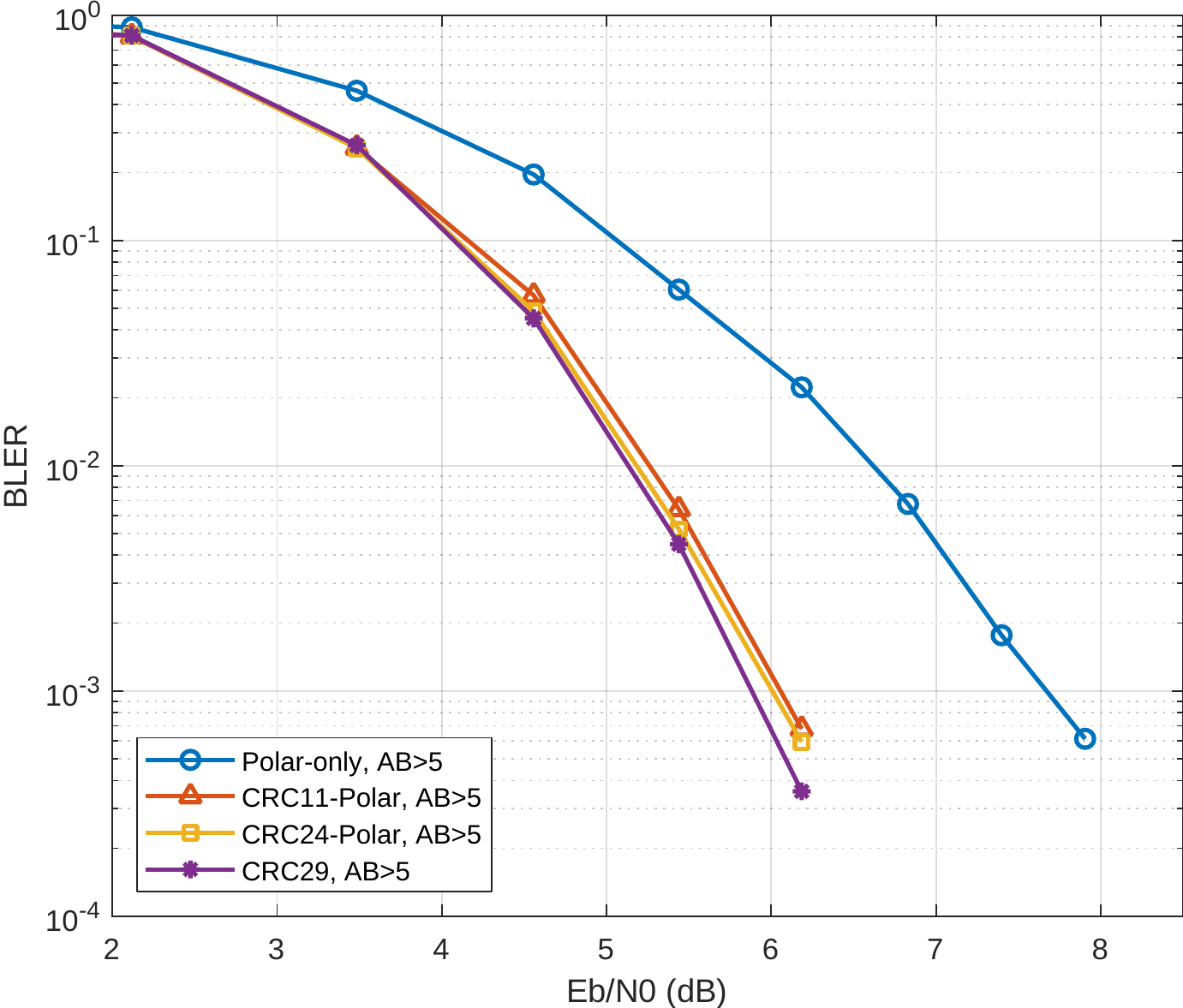}}
\caption{Hard detection performance of Polar(128,99), CA-Polar(128,99) with CRC11, CA-Polar(128,99) with CRC24 and CRC(128,99) (i.e. CRC29). All decoded with GRAND-SOS and $AB>5$.}
\label{fig:polar_crc}
\end{figure}

In 5G mobile systems, an 11-bit CRC (CRC11) is specified for uplink
CA-Polar codes, and a 24-bit CRC (CRC24) for downlink. We evaluate
performance of Polar codes with or without aid from CRC11 and CRC24,
and compare them to CRC codes from in Table \ref{tb:bch_crc}, 
for code-book with $(n,K)$ set to (128, 99). Hard detection ML performance when
decoded with GRAND-SOS are
presented in Fig. \ref{fig:polar_crc}. A generous abandonment
threshold $AB>5$ is chosen to avoid influence from computation
complexity on code performance. 

Let us first assess code performance under a hard detection scenario. The Polar(128, 99) code without the CRC shows significantly worse
performance than the three other codes. This is reasonable considering
that Polar codes are closely related to Reed-Muller codes
\cite{journals/tc/Muller54}, which are known to be inferior to cyclic
codes (such as BCH). The introduction of CRC codes, either CRC11
or CRC24, sufficiently breaks the undesirable structure in Polar
code, bringing its performance close to that achieved by CRC29.  In
this sense, the ``aid'' from CRC is so effective that it may be
preferable just to use the CRC as an error-correcting code, considering that significant
encoding complexity can be saved thanks to the simple implementation
of CRC encoders.

\begin{figure}[htbp]
\centerline{\includegraphics[width=0.44\textwidth]{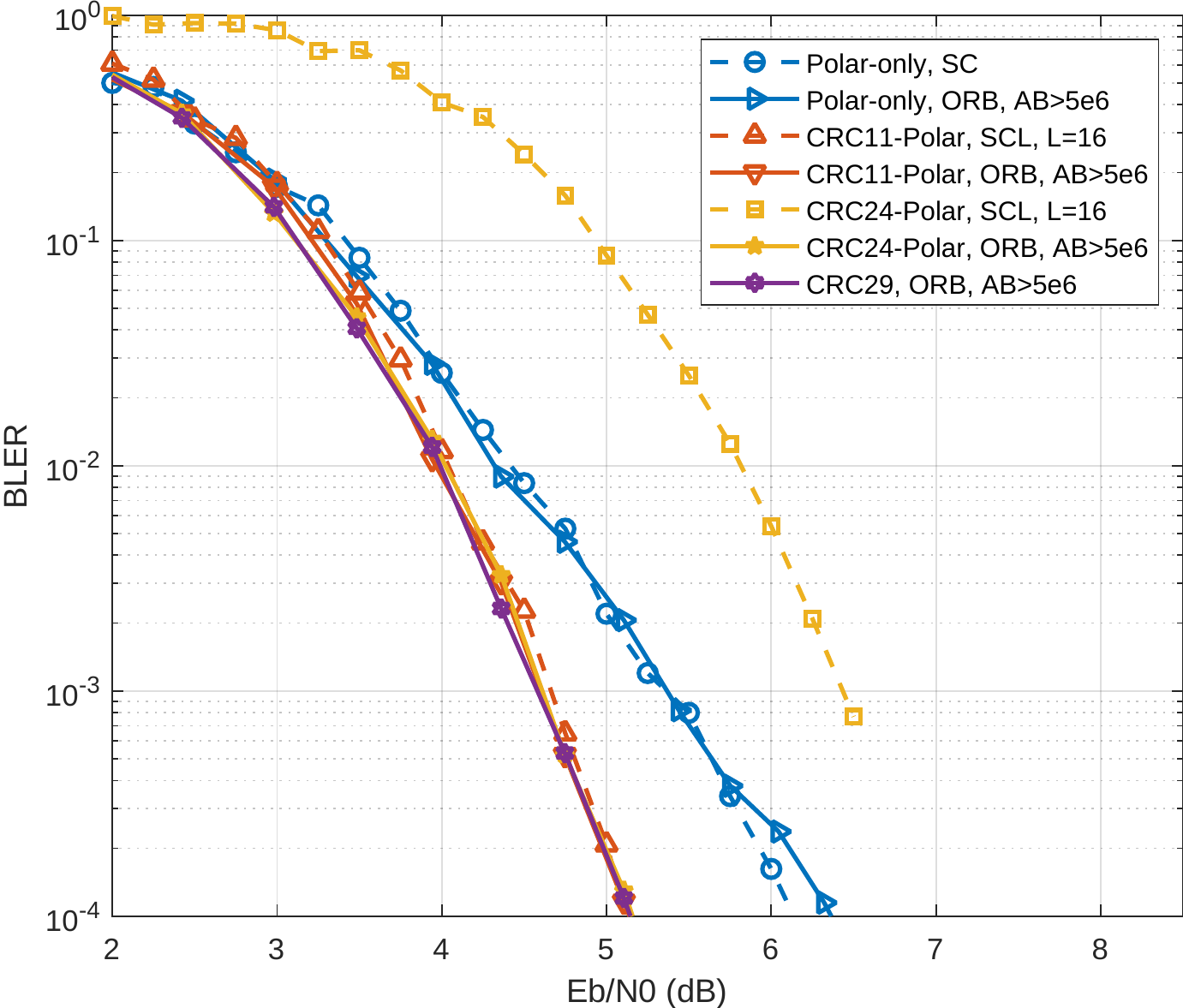}}
\caption{Soft detection performance of code-book setting of (128, 99) for Polar, CRC11 aided Polar, CRC24 aided Polar, and CRC29 codes; SC for Polar, CA-SCL with list size of 16 for CA-Polar and ORBGRAND with $AB>5 \times 10^6$ for all codes; SC and CA-SCL are performed with \cite{Cassagne2019a}.}
\label{fig:polar_soft}
\end{figure}

Having considered performance under hard detection, let us envisage
soft detection performance, which is of particular importance given that prevailing
decoders for Polar codes are soft decoders, especially when a
non-trivial decoding gain is expected \cite{lin2004error}. The extra gain comes from decoding errors beyond the hard detection decoding capability $t$. Therefore we use the number of code-book checks as the abandonment condition in this case, with $AB>5e6$ being a generous threshold that ensures optimal decoding performance. A gain 
of about 1.5 dB from soft decoding at BLER$=10^{-3}$ can be observed
by comparing Fig. \ref{fig:polar_crc} and Fig. \ref{fig:polar_soft}.
Of more interest is that under the same ORBGRAND decoding power
four candidate codes follow identical performance ranking as their
hard detection behaviors, i.e. CRC assists Polar codes to approach
the best performance achieved by the CRC code itself.

Evaluating the performance of the SC and CA-SCL soft decoding
algorithms in Fig. \ref{fig:polar_soft}, a surprising phenomenon
can be observed. While SC decoding of the Polar code and CA-SCL of
the CRC11 aided Polar code provide comparable performance to ORBGRAND,
the performance of CA-SCL with the CRC24 aided Polar code dramatically
degrades to a level even worse than the hard detection performance
of the same code in Fig. \ref{fig:polar_crc}. This reveals a previously
unnoticed issue with the CA-SCL algorithm. As a list decoding
algorithm, CA-SCL produces a set of candidate code-words in its
decoding procedure and use CRC checks to make candidate selection.
More CRC bits provide better error detection capability and
consequently higher confidence in candidate checks. However, the
redundancy introduced by CRC bits to the code-book is ignored for
error correction, leaving the remaining Polar bits to produce
the redundancy required for the error correcting list decoding.
With a given code-book  $(n,K)$ set of parameters, if the number of CRC bits is excessive relative to the
limited number of Polar redundant bits, then the decoder may experience limited space of list
candidates and, consequently,  performance loss, in spite of
the enhanced selection confidence from the extra CRC bits.  As the
key contributor to the error correcting quality of CA-Polar codes,
the CA-SCL algorithm has been applied mostly to long codes, in which
this shortcoming, which has been unnoticed so far to our knowledge, is far
from negligible on these short codes.

\begin{figure}[htbp]
\centerline{\includegraphics[width=0.44\textwidth]{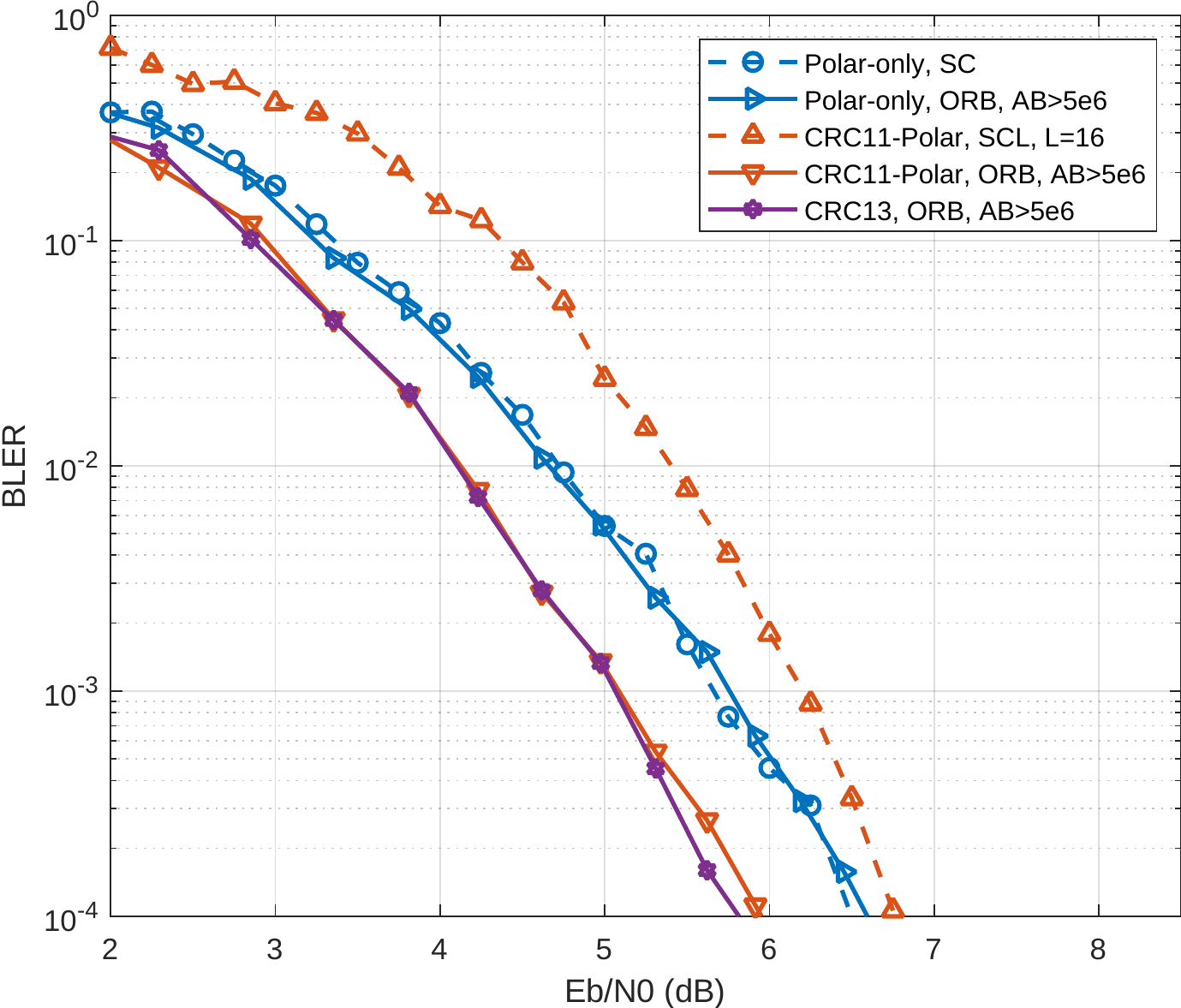}}
\caption{Soft detection performance of code-book setting of (64, 51) for Polar, CRC11 aided Polar, and CRC13 codes; SC for Polar, CA-SCL with list size of 16 for CA-Polar and ORBGRAND with $AB>5 \times 10^6$ for all codes; SC and CA-SCL are performed with \cite{Cassagne2019a}.}
\label{fig:polar_soft_64}
\end{figure}

In Fig. \ref{fig:polar_soft}, the CA-Polar (128, 99) with CRC11 
does not experience this performance loss, owing to the 18 Polar bits,
which are still sufficient to provide enough list candidate space.
Unfortunately this success is not guaranteed with shorter
code-word lengths or with higher coding rates. Fig.  \ref{fig:polar_soft_64}
displays the same short-coming of the CA-SCL algorithm with the
CRC11 but in CA-Polar(64,51), while ORBGRAND is stably successful.

The superior performance of CRC codes when compared to Polar codes,
CRC aided or not, with their ultra low encoding complexity, and the
robust decoding performance of ORBGRAND as compared to CA-SCL
algorithm, all suggest that a CRC code with ORBGRAND as its decoder
provides a better short code solution than Polar or CA-Polar codes.

\section{Summary} \label{sect:conclusion}
Despite being the most widely used error detection solution in
storage and communication, CRC codes have not been considered
seriously as error correcting codes owing to the absence of an
appropriate decoder. Instead, their main contribution has been in
assisting state-of-the-art list decoders, with CA-Polar codes as the
most successful recent example. With the invention of GRAND, the
possibility of considering CRCs as error correction codes can be
realized.

Here we consider hard and soft detection variants, GRAND-SOS and
ORBGRAND, to evaluate the performance of CRC codes in comparison
with other candidate short, high-rate codes, including BCH codes,
RLCs and CA-Polar codes. As short codes, CRCs not only provide at
least equivalent decoding performance to existing solutions, but
are superior in various aspects. Compared to to BCH codes, CRC codes
have no restriction on code lengths and rates. For very high rate
codes, RLCs are outperformed by CRC codes. Polar codes are surpassed
by CRC codes in both hard and soft detection BLER performance. Aided
by CRC, CA-Polar code can provide a matching performance, but the
CRC alone still wins with its ultra low encoding complexity. In
addition, CA-SCL decoder, as the key to practical adoption of
CA-Polar code, suffers significant performance loss when CRC bits
dominate the code's redundancy. The reason is that CA-SCL uses CRC
checks for candidate selection in list decoding, but does not avail
of the CRC bits for error correction. ORBGRAND, on the other hand,
is robust in all scenarios. Therefore, CRC with GRAND as its decoder
is a good solution to low latency applications requiring short
codes.

\bibliographystyle{IEEEtran}
\bibliography{my_bibli}

\end{document}